\documentclass[twocolumn,twocolumn]{IEEEtran}
\usepackage[T1]{fontenc}
\usepackage{color}
\usepackage{float}
\usepackage{mathrsfs}
\usepackage{amsmath}
\usepackage{amssymb}
\usepackage{graphicx}

\usepackage[unicode=true,
 bookmarks=true,bookmarksnumbered=true,bookmarksopen=true,bookmarksopenlevel=1,
 breaklinks=false,pdfborder={0 0 0},pdfborderstyle={},backref=false,colorlinks=false]
 {hyperref}
\hypersetup{pdftitle={Your Title},
 pdfauthor={Your Name},
 pdfpagelayout=OneColumn, pdfnewwindow=true, pdfstartview=XYZ, plainpages=false}

\makeatletter

\floatstyle{ruled}
\newfloat{algorithm}{tbp}{loa}
\providecommand{\algorithmname}{Algorithm}
\floatname{algorithm}{\protect\algorithmname}

\usepackage[caption=false,font=footnotesize]{subfig}
\usepackage{algorithm}
\usepackage{algorithmic}

\usepackage{multirow} %multirow for format of table 
\usepackage{amsmath} 
\usepackage{xcolor}

\allowdisplaybreaks[4]

\ifCLASSOPTIONcompsoc
\usepackage[caption=false,font=normalsize,labelfont=sf,textfont=sf]{subfig}
\else
\usepackage[caption=false,font=footnotesize]{subfig}
\fi

\usepackage{cite}
\usepackage{bm}
\usepackage{algorithmic}
\usepackage{algorithm}
\usepackage{graphicx}
\IEEEoverridecommandlockouts

\usepackage{lettrine}

\usepackage{geometry}
\@ifundefined{showcaptionsetup}{}{%
 \PassOptionsToPackage{caption=false}{subfig}}
\usepackage{subfig}
\makeatother

\begin{document}
\setlength{\parskip}{0em}
\bibliographystyle{IEEEtran}
\title{\textcolor{black}{Computation Offloading for Uncertain Marine Tasks by Cooperation of UAVs and Vessels}}

\author{\IEEEauthorblockN{Jiahao You$^{\dagger}$, Ziye Jia$^{\dagger}$, Chao Dong$^{\dagger}$, Lijun He$^{\ast}$, Yilu Cao$^{\dagger}$, and Qihui Wu$^{\dagger}$\\}
\IEEEauthorblockA{$^{\dagger}$The Key Laboratory of Dynamic Cognitive System of Electromagnetic Spectrum Space, Ministry of Industry and Information Technology, Nanjing University of Aeronautics and Astronautics, Nanjing, Jiangsu, 210016, China\\
$^{\ast}$School of software, Northwestern Polytechnical University, Xi'an, Shaanxi, 710129, China\\
\{yjiahao, jiaziye, dch, caoyilu, wuqihui\}@nuaa.edu.cn, lijunhe@nwpu.edu.cn}}

\maketitle
\pagestyle{empty} 
\thispagestyle{empty}

\begin{abstract}
With the continuous increment of maritime applications, the development of marine networks for data offloading becomes necessary. 
However, the limited maritime network resources are very difficult to satisfy real-time demands. 
Besides, how to effectively handle multiple compute-intensive tasks becomes another intractable issue. 
Hence, in this paper, we focus on the decision of maritime task offloading by the cooperation of unmanned aerial vehicles (UAVs) and vessels.
Specifically, we first propose a cooperative offloading framework, including the demands from marine Internet of Things (MIoTs) devices and resource providers from UAVs and vessels. 
Due to the limited energy and computation ability of UAVs, it is necessary to help better apply the vessels to computation offloading.
Then, we formulate the studied problem into a Markov decision process, aiming to minimize the total execution time and energy cost.
Then, we leverage Lyapunov optimization to convert the long-term constraints of the total execution time and energy cost into their short-term constraints, further yielding a set of per-time-slot optimization problems.
Furthermore, we propose a Q-learning based approach to solve the short-term problem efficiently.
Finally, simulation results are conducted to verify the correctness and effectiveness of the proposed algorithm.
\end{abstract}

\begin{IEEEkeywords}
Maritime networks, Cooperative computation offloading framework, Lyapunov optimization, Q-learning.
\end{IEEEkeywords} 

\section{Introduction}\label{s1}
\IEEEPARstart{W}{ith} the increasing demands from marine Internet of things (MIoTs) \cite{Zhu_Software_2022}, the limited marine network resources cannot satisfy multiple tasks. Fortunately, the development of space-air-ground network can help figure out such resource insufficient issues. Besides, the multi-access edge computing (MEC) \cite{Jia-Hierarchical_2022,dong_uavs_2021,Jia_LEO_2021} technique can be introduced to handle the computation related requirements from MIoTs, e.g. marine rescue. 
Due to the easy deployment and flexible movement of unmanned aerial vehicles (UAVs), it can be employed to assist MIoTs for rapid rescue \cite{wang_hybrid_2021}. 
However, UAVs cannot complete the compute-intensive tasks due to the limited energy and computation resources. 
Therefore, how to assist UAVs to handle compute-intensive tasks is significant. 
Vessels can provide abundant computing and energy resources to address the resource limitation of UAVs. 
Thus, the cooperation of UAVs and vessels provides more flexible services for the maritime MEC tasks.
However, the following challenges still should be handled: 
1) the position adjustment of UAV cannot follow the uncertainty  of user demands;
2) how to select an appropriate vessel for a UAV to offload the selected tasks;
3) the arrival of marine tasks is uncertain, and UAVs have no knowledge of the task information.
Hence, it is significant to design a reasonable offloading scheme involving both UAVs and vessels for the marine MEC tasks.

There exist a couple of recent works related to MEC in UAV based scenarios. 
For instance, 
Jia \textit{et al.} \cite{Jia_Joint_HAP_2021} consider the cooperation of LEO satellites and HAPs. The matching among users, HAPs, and satellites are solved with a matching algorithm.
In \cite{yang_multi-armed_2022}, Yang \textit{et al.} use multi-armed bandits learning to optimize offloading delay and energy cost. Thus, the reliability of marine communications is ensured.
Zhao \textit{et al.} \cite{zhao_multi-agent_2022} study cooperative offloading strategies under MEC for multiple UAVs, and model the problem as a Markov decision process (MDP). The resource management of computation and communication is implemented by multi-agent deep reinforcement learning to minimize the sum of latency and energy cost.
Ning \textit{et al.} \cite{ning_5g-enabled_2021} build a 5G-enabled UAV-to-community offloading system to maximize the throughput. This work designs an auction based algorithm to deal with the trajectory design, as well as a community based scheduling algorithm to meet the transmission rate demands.
In \cite{sacco_sustainable_2021}, Sacco \textit{et al.} solve the problem of task offloading via a distributed architecture. The system learns the best actions from environments to minimize the latency and energy cost to improve network service levels. However, the aforementioned works lack considering the uncertain MEC tasks in the marine environment, which is a significant issue.

In this work, we focus on the MEC related problem cooperated by UAVs and vessels for marine applications. In detail,  
we propose an integrated space-air-ground-marine architecture in order to clearly depict the resources and demands in the marine scenario. 
Based on the proposed architecture, we further formulate the studied problem into a mixed-integer program, to minimize both the total execution time and energy cost. 
However, the formulated problem is NP hard and prone to dimensional catastrophes at large scales \cite{Shuai_Transfer_2022}. 
To this end, we first utilize the Lyapunov optimization technique to transform the formulation problem into a set of per-time-slot optimization problems.
Then, we propose a Q-learning based method to solve each per-time-slot optimization problem efficiently.
Finally, numerical results verify that the proposed algorithm can effectively guarantee solution effectiveness.

The rest of the paper is organized as follows. The system model and problem formulation are presented in Section \ref{s2}. Then, a virtual queue-based Q-learning approach (VQQ) is proposed in Section \ref{s3}. Numerical results are provided in Section \ref{s4}, and finally conclusions are drawn in Section \ref{s5}.

\section{SYSTEM MODEL AND PROBLEM FORMULATION}\label{s2}
\begin{figure}[!t]
  \centering
  \includegraphics[width=8cm]{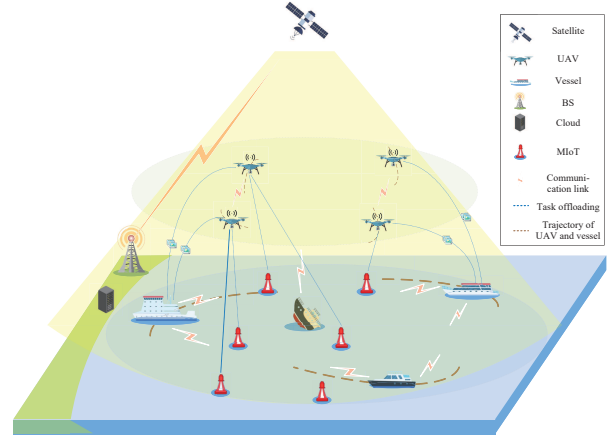}
  \caption{System overview.}
  \label{f1}
\end{figure}

As shown in Fig. \ref{f1}, an integrated space-air-ground-marine architecture consisting of satellites, UAVs, vessels, and base stations (BSs) is provided. Satellites provide signaling and navigation for the target area. BSs mainly provide tasks commands for vessels and UAVs. The vessels are equipped with multiple receiving antennas, powerful computing capability, etc. Hence, vessels can perform as the mobile base stations. The MIoT devices in Fig \ref{f1} build multiple tasks with computation demands. However, the limited computing and energy ability of MIoT cannot complete the local computation. Therefore, UAVs can collect the data from MIoT for lightweight computing. Besides, due to the limited payloads of UAVs, the compute-intensive demands can be relayed to vessels by the UAV. 

The components in the system include MIoTs, UAVs, and vessels. MIoTs are denoted as $i\in\mathcal{M}=\{1 , 2, 3, ... , I\}$, UAVs are represented as $j\in\mathcal{U}=\{1, 2, 3,... , J\}$, and vessels are indicated by $k\in\mathcal{V}=\{1, 2, 3,... , K\}$. The vessels are equipped with $N_k^{max}$ antennas, denoting the maximum number of UAVs that can communicate simultaneously. 
The total time is depicted as $\mathcal{T}$ and divided into $T$ time periods with length $l_0$ of each time slot. We utilize a 3D Cartesian coordinate system to represent the positions of MIoTs, UAVs, and vessels, as $m_i(t) = (x_i^{m}, y_i^{m}, h_i^m)$, $u_j(t) = (x_j^{u}, y_j^{u},h_j^u)$, and $v_k(t) = (x_k^{v}, y_k^{v}, h_k^v)$, respectively. The tasks generated by the MIoTs in time slot $t$ is $\lambda_i^{m}(t) = \{l_i^m(t), d_i^m(t)\}$,  in which, $l_i^m(t)$ (in bits) indicates the input data size, and $d_i^m(t)$ (in CPU cycles/bit) denotes the number of CPU cycles to process data.
Considering the practical situation of limited UAV capacity, the maximum capacity of UAV $j$ is denoted as $S_j^{max}$.

In time slot $t$, the task offloading decision should be made for each new arrival task, and we define the task offloading decision $\mathcal{O}=[o_{i,j}]_{I\times J}$, $\mathcal{S} =[s_{i,k}]_{J\times K}$ as:
\begin{equation}\label{e1}
o_{i,j}(t)= \begin{cases}1,&\text { if task }  \lambda_i^{m}(t) \text { is offloaded to UAV } j, \\ 
0,&\text { otherwise,}\end{cases}
\end{equation}
and
\begin{equation}\label{e2}
s_{i,k}(t)= \begin{cases}1, & \text { if task } \lambda_i^{m}(t) \text { is offloaded to vessel } k, \\ 0,&\text { otherwise. }\end{cases}
\end{equation}
The connection decision $\mathcal{Q}=[q_{i,j}]_{I\times J}$ between MIoT $i$ and UAV $j$ is:
\begin{equation}\label{e3}
q_{i,j}(t)= \begin{cases}1, & \text {if MIoT } i \text { is connected to UAV } j, \\ 0, & \text { otherwise. }\end{cases}
\end{equation}
Besides, some tasks should be offloaded to vessels relayed by UAVs, the connection $\mathcal{P}=[p_{j,k}]_{J\times K}$ between UAV $j$ and vessel $k$ is defined as:
\begin{equation}\label{e4}
p_{j,k}(t)= \begin{cases}1, & \text { if UAV } j  \text { is connected to vessel } k, \\ 0, & \text { otherwise. }\end{cases}
\end{equation}
%符号描述位置

Note that we assume UAVs communicate in full-duplex mode, i.e., the upload and download of tasks can be performed simultaneously. 
Moreover, since the size of computed data is much smaller than the original data, the orthogonal frequency division multiple access technique is leveraged in backhaul transmission.
Hence, the result can be sent back within one time slot, and the corresponding delay is omitted.

\begin{figure*}[htbp]
  \centering
  \subfloat[Tasks are generated from MIoTs.]{
  \includegraphics[width=8.5cm]{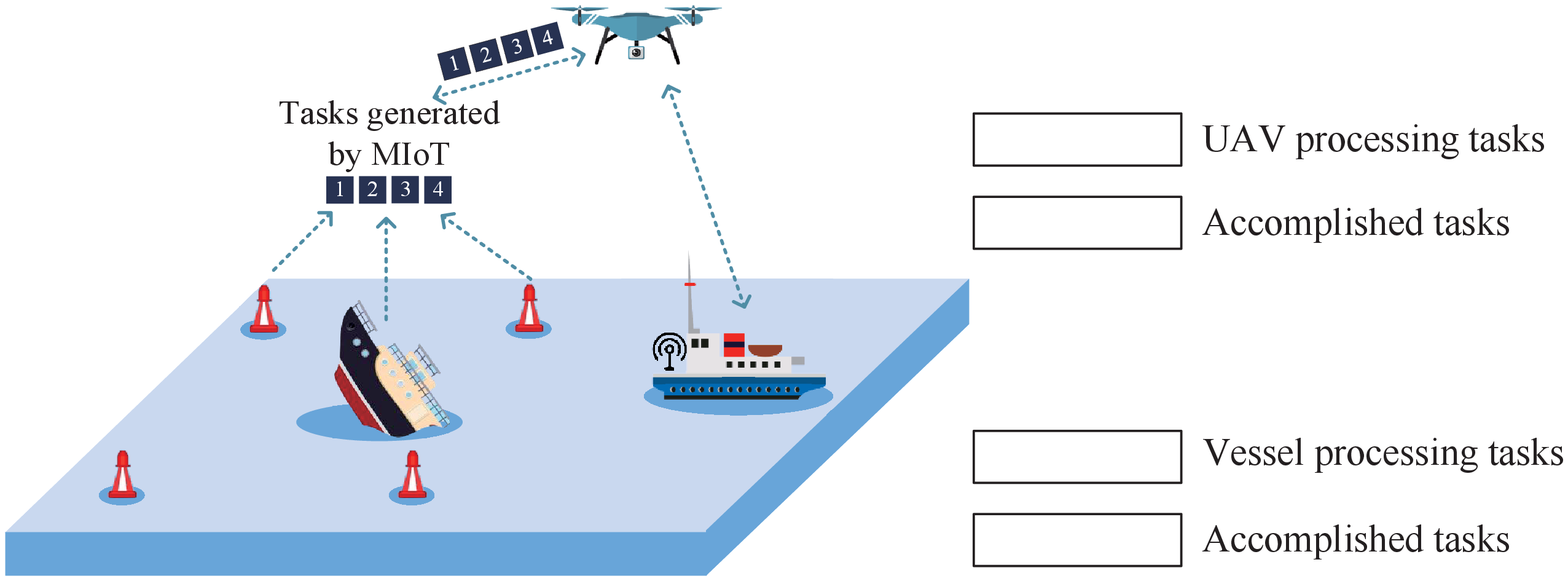}
  }
  \quad
  \subfloat[Task 1 is being processed on the UAV. Task 2 is offloaded to the vessel for processing.]{
  \includegraphics[width=8.5cm]{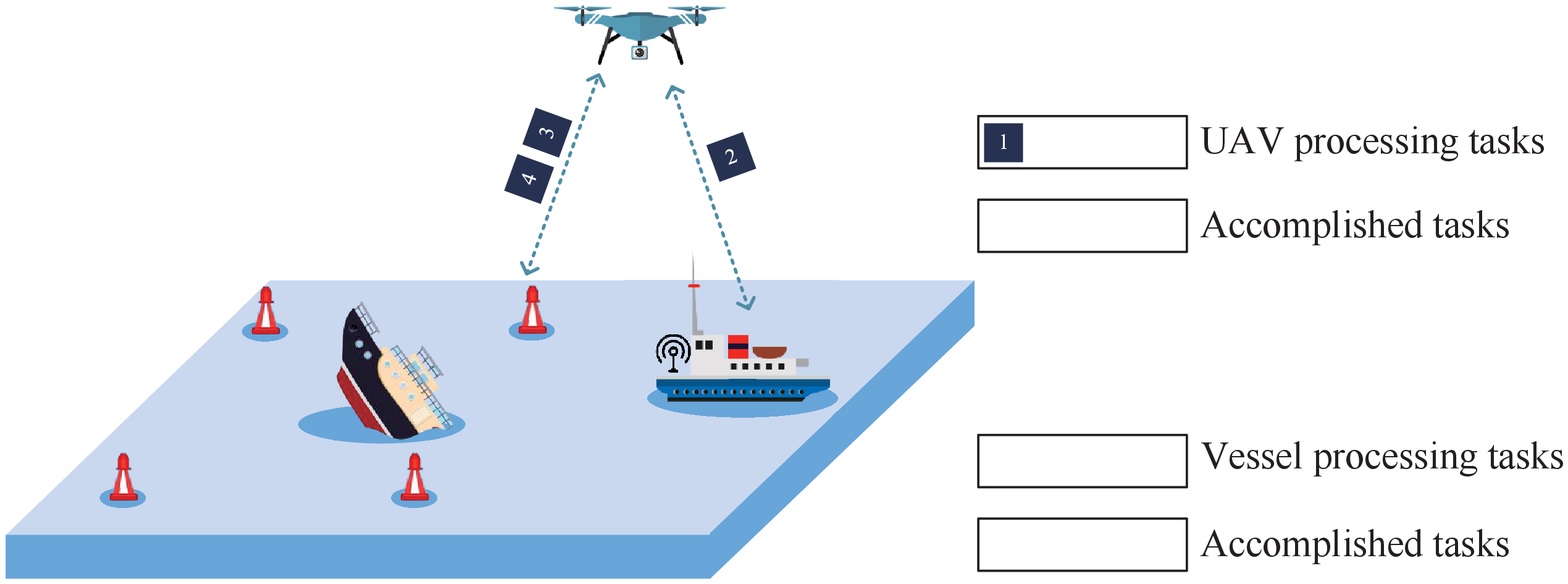}
  }
  \quad
  \subfloat[Task 1 is completed and transmitted back to MIoT. Task 2 is being processed on the UAV. Task 3 is offloaded to the vessel for processing. Task 4 is being processed on the UAV.]{
  \includegraphics[width=8.5cm]{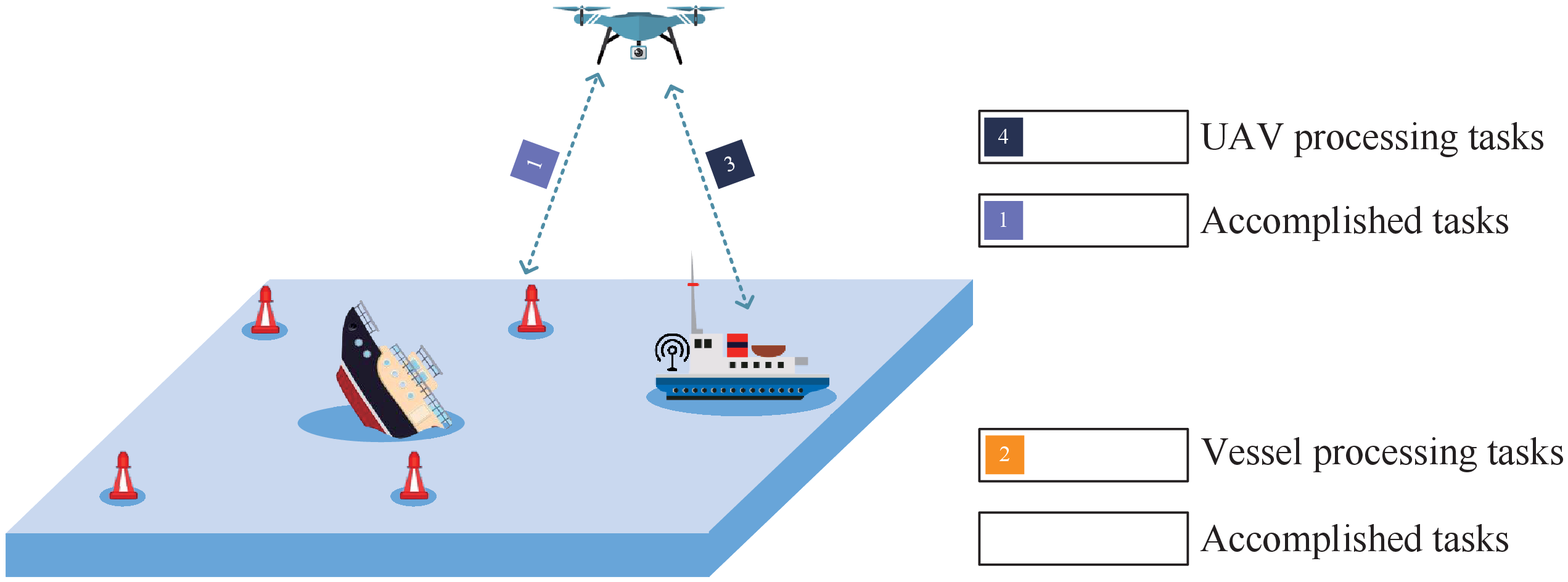}
  }
  \quad
  \subfloat[Task 2 is completed and transmitted back to UAV. Task 3 is being processed on the vessel. Task 4 completion is returned to the MIoT.]{
  \includegraphics[width=8.5cm]{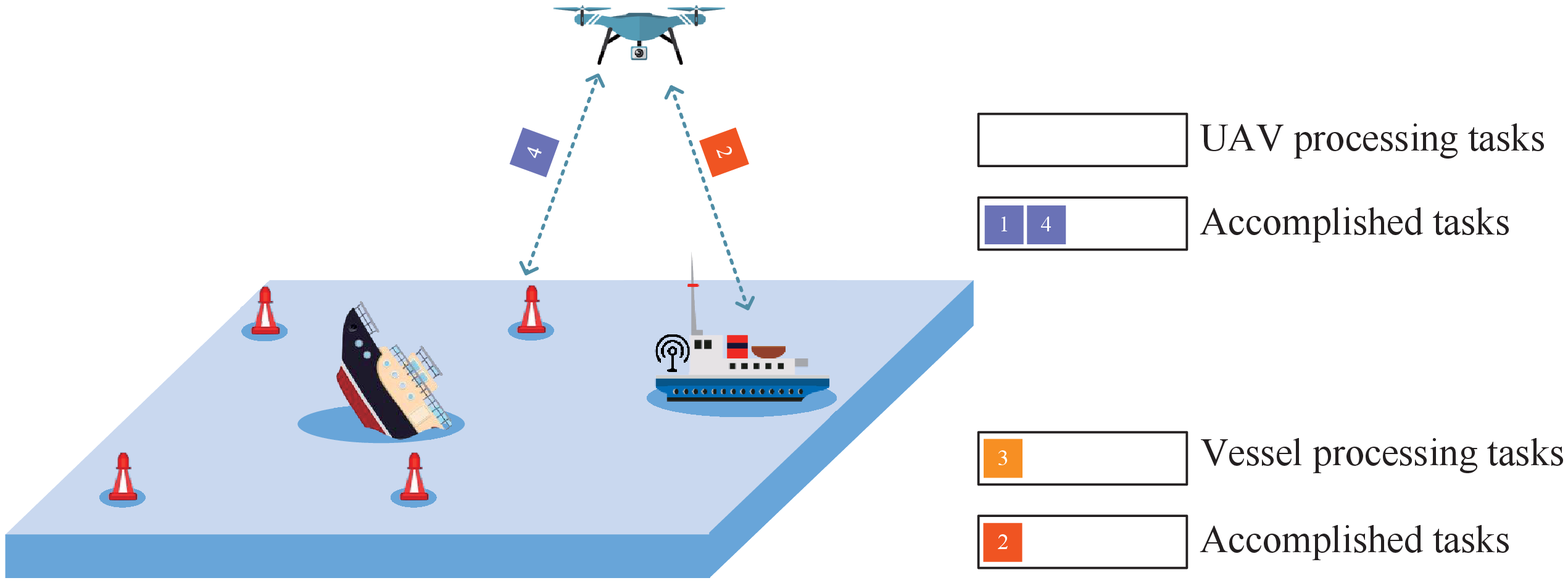}
  }
  \caption{An example of task offloading}
\end{figure*}

\subsection{Communication Model}
According to \cite{wang_hybrid_2021}, $\xi_{a,b}(t)$ indicates the path loss from node $a$ to node $b$, in which ($a \in \mathcal{M}$, $b \in\mathcal{U}$) or ($a\in \mathcal{U}$, $b \in \mathcal{V}$).
The path loss between two nodes in the marine environment is:
\begin{equation}
\begin{aligned}
\xi_{a,b}(t)&=\frac{\zeta_{L}-\zeta_{NL} }{1+\alpha  \exp \{ -\beta [\gamma_{a,b}(t) -\alpha \} }\\
&+20 \lg\left(\frac{4\pi\|x_a-x_b\| f_c}{C_0}\right)+\zeta_{NL},
\end{aligned}
\end{equation}
in which
\begin{subequations}
\begin{flalign}
  & \gamma_{a,b}(t) = \arctan \left(\frac{\vert h_a-h_b \vert }{\|x_a-x_b\|_2 }\right),
\end{flalign}
\end{subequations}
where $x_a,x_b$ denote the positions of transmitter $a$ and receiver $b$, respectively.  $h_a$ and $h_b$ represent the heights of transmitter and receiver, respectively. $f_c$ indicates the carrier frequency, $C_0$ is the speed of light. $\zeta_{L}$, $\zeta_{NL}$, $\alpha$, and $\beta$ represent environmental parameters.

Following Shannon formula, the average rate between transmitter $a$ and receiver $b$ is calculated as:

\begin{equation}
R_{a,b}(t)=B_0 \log_2  [1+\frac{P_{a}\xi_{a,b}(t)}{N_G}],
\end{equation}
where $B_0$ is the bandwidth of the channel, $P_{a}$ represents the transmitting power, and $N_G$ indicates the power of the additive white Gaussian noise.

\subsection{Computation Model}
\subsubsection{UAV based computation model}
For each task generated by MIoT $i$ in time slot $t$, 
the total execution time includes the execution time of the UAV-based computation, the queuing delay and the task transmission time:
\begin{equation}
\begin{aligned}
T_{i,j}^{uc}(t)&=T_{i,j}^{ue}+T_{i,j}^{ut}+T_{i, j}^{uqd}\\
            &=\frac{d_i^{m}(t)}{f_j^u}+\frac{l_i^m(t)}{R_{i,j}(t)}+T_{i, j}^{vqd},
\end{aligned}
\end{equation}
where $T_{i,j}^{ue}$ is the execution time of UAV-based computation. 
$T_{i,j}^{ut}$ represents the transmission time from MIoT $i$ to UAV $j$, and $T_{i, j}^{uqd}$ denotes the queuing delay. $f_j^u$ indicates the computing ability of UAV $j$, i.e., the CPU clock.

\subsubsection{Vessel based computation model}
If a task is offloaded from UAV $j$ to vessel $k$, such a process requires three procedures: task transmission, task execution, and task backhaul. Since the backhaul time is much smaller than the transmission and execution time, it is omitted and the total execution time is expressed as:
\begin{equation}
  \begin{aligned}
    T_{i,j,k}^{vc}&=T_{j,k}^{ve}+T_{i,j}^{ut}+T_{j,k}^{vt}+T_{j,k}^{vqd}\\ 
                &=\frac{d_i^m(t)}{f_k^v}+\frac{l_i^m(t)}{R_{i,j}(t)}+\frac{l_i^m(t)}{R_{j,k}(t)}+T_{i,k}^{vqd},
  \end{aligned}
\end{equation}
where $T_{j,k}^{ve}$ is the computation time cost on vessel $k$, and $T_{j,k}^{vt}$ denotes the task transmission time from UAV $j$ to vessel $k$. $f_k^v$ represents the computing ability of vessel $k$, i.e., the CPU clock.

As above, the total execution time for task generated by MIoT $i$ in time slot $t$ is:
\begin{equation}
T_i^a(t) = \sum_{j=1}^{J}\sum_{k=1}^{K}o_{i,j}(t)T_{i,j}^{uc}(t)+s_{i,k}(t)T_{i,j,k}^{vc}.
\end{equation}

During task offloading, the constraints related to the total execution time should be considered, to ensure the effectiveness. 
Hence, the total execution time related constraint is listed as:
\begin{equation}\label{e11}
  \lim _{T \rightarrow \infty} \frac{1}{T} \sum_{t=1}^T \frac{T_i^a(t)}{\overline{T_i^a}(t)} \leq \varphi_{max}^t, \\
\end{equation}
where $\overline{T_i^a}(t) = \frac{1}{t}\sum_{i=0}^{t-1}T_i^a(t)$, and $\varphi_{max}^t$ is the bounds of the total execution time.

\subsection{Energy Model}
There exist three main energy-consuming modes of UAVs: hovering, trajectory and for computation functions.
The energy required for UAV $j$ trajectory is \cite{al-habob_energy-efficient_2022}:
\begin{equation}
P_j^t=\frac{v_j}{v_j^{\max}}\left[P_j^{\max }-P_j^h\right],
\end{equation}
where $v_j$ and $v_j^{max}$ denote the current speed and maximum speed, respectively. $P_j^{\max}$ is the power at maximum speed and $P_j^{h}$ is the power for hovering. 
The hovering energy cost of the UAV is:
\begin{equation}
P_j^{h}=C_0\sqrt{\frac{\left(M_j^u\right)^3}{r_j^2 \kappa_j}},
\end{equation}

where $g$ represents the acceleration of gravity, and $\theta$ denotes the air density.
$M_j^u$ represents the mass, $r_j$ is the propeller radius, and $\kappa_j$ indicates the number of propellers. 
$C_0$ is the environmental parameter, and $C_0=\sqrt{\frac{g^3}{2 \pi \theta}}$.

The energy cost of a UAV to compute the task generated by MIoT $i$ at time slot $t$ is:
\begin{equation}
E_{i,j}^{uc}(t) = d_i^{m}(t)\varsigma_j^u,
\end{equation}
where $\varsigma_j^u$ is the energy cost per unit computational resources. The energy consumed by vessel $k$ to compute task $\lambda_i^{m}(t)$ is
\begin{equation}
E_{i,j,k}^{vc} = d_i^{m}(t)\varsigma_k^v,
\end{equation}
where $\varsigma_k^v$ is the energy cost of per unit of computational resources of vessel $k$. 

Hence, the energy cost of the task generated by MIoT $i$ in time slot $t$ is:
\begin{equation}
\begin{aligned}
  E_i^a(t)  &= \underbrace{\sum_{j=1}^{J}o_{i,j}(t)P_j^t\frac{\Vert u_j(t)-u_j(t+1) \Vert}{v_j}+P_j^hl_0}_{\text{energy cost for UAV movement}}\\
  &+\underbrace{ \sum_{j=1}^{J}\sum_{k=1}^{K}o_{i,j}(t)E_{i,j}^{uc}+\frac{s_{i,k}(t)P_jL_i^m(t)}{R_{j,k}(t)}}_{\text{UAV based computation and communication cost}}\\
  &+\underbrace{\sum_{j=1}^{J}\sum_{k=1}^{K}s_{i,k}(t)E_{i,j,k}^{vc}}_{\text{vessel based calculation cost}}.
\end{aligned}
\end{equation}

% During task offloading, the constraints related to the total execution time and energy should be considered, to ensure the effectiveness and energy efficiency. 
% Hence, the total execution time and task energy related constrained respectively listed as:
% \begin{equation}
%   \lim _{T \rightarrow \infty} \frac{1}{T} \sum_{t=1}^T \frac{T_i^a(t)}{\overline{T_i^a}(t)} \leq \varphi_{max}^t, \\
% \end{equation}
Besides, energy related constraints should be considered in terms of time constraints. Therefore, task energy related constrained is represented as:
\begin{equation}\label{e17}
  \lim _{T \rightarrow \infty} \frac{1}{T} \sum_{t=1}^T \frac{E_i^a(t)}{\overline{E_i^a}(t)} \leq \varphi _{max}^e,
\end{equation}
where $\overline{E_i^a}(t) = \frac{1}{t}\sum_{i=0}^{t-1}E_i^a(t)$, and $\varphi _{max}^e$ is the bounds of energy constraints.

\subsection{Problem Formulation}
The optimization problem is formulated to minimize the total execution time as well as energy cost:
\begin{subequations}
\begin{align}
\textbf{P0:}\min_{\mathcal{O},\mathcal{S},\mathcal{P},\mathcal{Q}}&\sum_{t=1}^T\sum_{i=1}^{I}\theta_0 E_i^a(t)+\theta_1 T_i^a(t), \label{Za}\\
\text{s.t. } &\sum_{j=1}^{J}o_{i,j}(t)+\sum_{k=1}^K s_{i,k}(t)= 1,\forall i,t,\label{Zb}\\
&\sum_{i=1}^I o_{i,j}(t)*l_i^m(t) + S_j(t) 	\leq S_j^{max},\forall j,t,\label{Zc}\\
&\sum_{j=1}^J p_{j,k}(t)\leq N_k^{max},\forall k,t,\label{Zd}\\
&\|u_j(t)\!-\!u_j(t\!+\!1)\| \!\leq\! l_0V_j^{max}\!,\!\forall t \in [0,T\!-\!1],j,\label{Ze}\\
&o_{i,j}(t),s_{i,k}(t),q_{i,j}(t),p_{j,k}(t) \in\{0,1\},\forall i,j,k,t,\label{Zf}\\
& (\ref{e11})\ \text{and}\ (\ref{e17}),\forall i,j,k,\notag
\end{align}
\end{subequations}
where $\theta_0$ and $\theta_1$ indicate the weights of energy cost and the total execution time, respectively.
Constraint (\ref{Zb}) ensures that a task can be implemented by only one UAV or vessel. 
Constraint (\ref{Zc}) is a storage constraint, denoting at any time slot, a UAV cannot store more tasks than its own storage capacity. 
Constraint (\ref{Zd}) denotes a vessel connection constraint i.e., the maximum number of connections of a vessel cannot exceed $N_k^{max}$. 
Constraint (\ref{Ze}) indicates that the maximum moving distance for a UAV in two adjacent time slots is limited by the speed of UAV.
Constraint (\ref{Zf}) indicates the binary indicators. 
Due to the binary variables, P0 is non-convex and the solution space grows exponentially with the problem scale, which is intractable to solve by a traditional optimization approach.

\section{DEEP REINFORCEMENT LEARNING-BASED TASK OFFLOADING}\label{s3}
Since both the constraints of energy cost and time cost hold for a long time $\mathcal{T}$, i.e., long-term constraints. Hence, $\mathcal{T}$ is decoupled as multiple time slots, and each time slot is short-term $t$.
Then, by leveraging the Lyapunov optimization, the long-term stochastic optimization problem can be decoupled into short-term deterministic optimization subproblems.

\subsection{Problem Transformation}
In order to transform the constraints in P0 within the same time slot, we convert (\ref{e11}) and (\ref{e17}) into virtual queues, detailed as:
\begin{equation}
V_i^t(t+1) =  \max\{V_i^t(t)+\frac{T_i^a(t)}{\overline{T_i^a}(t)}-\varphi _{max}^t,0\},\\
\end{equation}
and
\begin{equation}
V_i^e(t+1) =  \max\{V_i^e(t)+\frac{E_i^a(t)}{\overline{E_i^a}(t)}-\varphi _{max}^e,0\}.
\end{equation}
Thus, the original problem P0 is converted to the cross-time slot problem  P1. 
Therefore, P1 can be solved by per-time-slot strategy, which guarantee the time and energy constraints:
\begin{equation}\label{e22}
\begin{aligned}
  \textbf{P1:}&\min_{\mathcal{O}(t),\mathcal{S}(t),\mathcal{Q}(t),\mathcal{P}(t)}\sum_{t=1}^T\sum_{i=1}^{I}\mathfrak{A}_{i,j,k}(t),\\
  &\text{s.t. } 18(b)-18(f),\\
\end{aligned}
\end{equation}
where $\mathfrak{A}$ is:
\begin{equation}
\begin{aligned}
\mathfrak{A}_{i,j,k}(t) = V_i^t(t)(\frac{T_i^a(t)}{\overline{T_i^a}(t)}-\varphi_{max}^t)+V_i^e(t)(\frac{E_i^a(t)}{\overline{E_i^a}(t)}-\varphi_{max}^e).
\end{aligned}
\end{equation}

To figure out P1, it is necessary to design a task offloading scheme, and each MIoT generates a task in different time slots and offloads to a UAV or a vessel, to minimize $\mathfrak{A}_{i,j,k}(t)$. Hence, P1 is transformed as a MDP problem \cite{zhu_learning-based_2021}.

\begin{algorithm}[t]
\caption{A virtual queue-based Q-learning approach.}
\begin{algorithmic}[1] %这个1 表示每一行都显示数字
\REQUIRE  State $S(t)$, epochs $N$, and $\epsilon$.
\ENSURE  $\mathcal{O}(t)$, $\mathcal{S}(t)$, $\mathcal{Q}(t)$, and $\mathcal{P}(t)$.
\STATE Randomly generate initial values of $s(t)$ and $\mu$, and create neural networks with $\mu$.
\FOR{each $k = 1:N$}
\FOR{each $t = 1:T$}
\IF {rand$(0, 1)< \epsilon$}
\STATE Stochastically select action.
\ELSE
\STATE Select the action with the largest Q value.
\ENDIF
\STATE Change the connections according to $\mathcal{O}(t)$ and $\mathcal{S}(t)$. 
\STATE Offload tasks according to $\mathcal{Q}(t)$ and $\mathcal{P}(t)$.
\STATE Compute the reward of the task and loss function.
\STATE Update $\mu = \mu+\phi\triangledown_{\mu}loss(\mu)$.
\STATE Update $s(t+1)$ and move to the next state.
\ENDFOR
\ENDFOR
\end{algorithmic}
\end{algorithm}

In particular, the process of MDP is defined as: state, action, reward and transfer probability, as follows:
\begin{enumerate}
  \item State: The state at time slot $t$ is defined as $s(t)=\{T_i^a(t),   E_i^a(t), V_i^t(t), V_e^t(t)\}$, which represents the energy cost and time consumption in the current time slot.
  \item Action: The action in time slot $t$ is defined as $a(t)=\{\mathcal{O}(t),\mathcal{S}(t),\mathcal{Q}(t),\mathcal{P}(t)\}$, which denotes the offloading and connection decision.
  \item Reward: As for the optimization problem, we set the reward as $r(t)=-\mathfrak{A}_{i,j,k}(t)$.
  \item Transfer probability: The probability of moving from the current state to the next state is $p(s(t+1)\mid s(t), a(t))$.
\end{enumerate}

In MDP, the state space solution tends to deteriorate due to the curse of dimensionality. Then, we propose the reinforcement learning-based model to figure out the intractable issue.

\subsection{Algorithm Design}
The Q-learning based mechanism is utilized to solve the current problem via selecting the Q value of action $a(t)$ in state $s(t)$, and the Q value is defined as:
\begin{equation}
\begin{aligned}
    Q(s(t), a(t))=& Q(s(t), a(t))+\phi[r({t+1})+\\
    &\psi \max _a Q(s({t+1}), a)-Q(s(t), a(t))],
    \end{aligned}
\end{equation}
where $\phi, R,\psi$ denote the learning rate, reward and discount factors,  respectively.

In order to simplify the scale of the computation and reduce the subsequent storage of Q values, we design the deep Q-network method, leveraging a deep neural network to map the relationship between the current state and the action. Such a relationship denotes Q values, and the deep neural network is utilized to gradually approximate the current parameter $\mu$:
\begin{equation}
Q(s(t),a(t),\mu)\approx Q_{\mathcal{O},\mathcal{S},\mathcal{P},\mathcal{Q}}(s(t),a(t)).
\end{equation}
Further, the loss function is defined as:
\begin{equation}\label{e25}
loss(\mu)= (r(t)+\psi \max _a Q(s({t+1}), a)-Q(s(t), a(t)))^2.
\end{equation}

The VQQ is detailed in Algorithm 1. Firstly, we assume that there are $N$ epochs and each episode consists of $T$ time slots. At the beginning of each time slot, the offloading policy and linking policy are chosen according to the random probability $\epsilon-$greedy (line 4-line 7). Then, the tasks are offloaded to the selected UAV or vessel according to the action (line 9-line 10). Based on (\ref{e22}) and (\ref{e25}), the reward and loss function are calculated for this time slot $t$ (line 11). Finally, the value of the next state is updated (line 12-line 13).

\section{SIMULATIONS AND PERFORMANCE ANALYSIS}\label{s4}
\begin{figure}[!t]
  \centering
  \includegraphics[width=8cm]{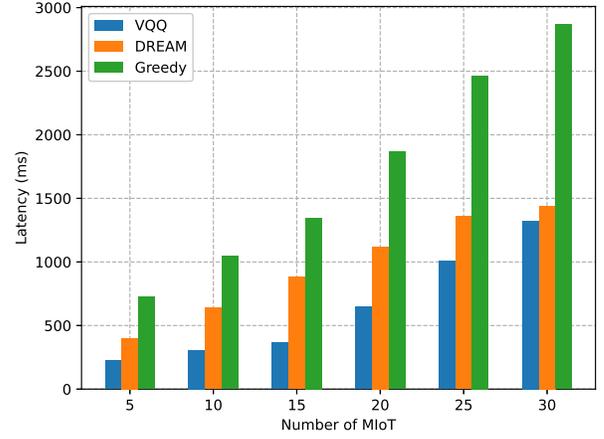}
  \caption{Effect of number of MIoT on latency.}
\end{figure}

\begin{figure}[!t]
  \centering
  \includegraphics[width=8cm]{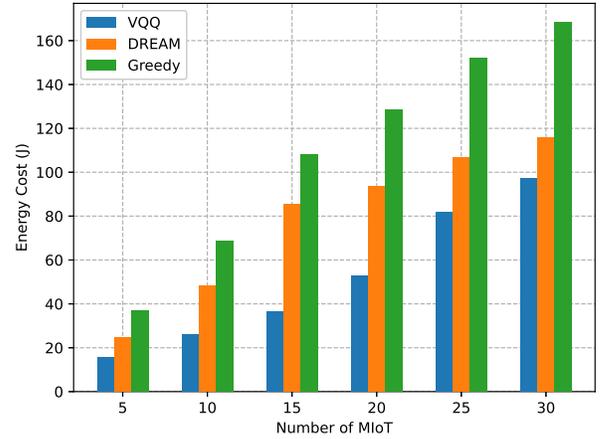}
  \caption{Effect of number of MIoT on time cost.}
\end{figure}
Simulations are conducted in the following scenario: 3 vessels, 10 UAVs and $I$ MIoTs are evenly distributed in a region of 5km$\times$5km. The vessels and MIoTs heights are set as 0 and the UAV's height set is 100m. Python is employed to implement the proposed algorithm VQQ. In addition, other parameters are set as: $T=$ 100, $l_0=$ 1s, $l_i^m(t) \in$ [2.4, 2.6]Mbits, $d_i^m(t)\in$ [2.4, 3.6]Mbits, $B_0 =$ 1Mhz, $\zeta_{L}=$ 2.3, $\zeta_{NL}=$ 34, $\alpha=$ 5.0188, $\beta = $ 0.3511, $P_{a} =$ 10W, $N_G =$ -114dBm, $f_c =$ 1Mhz, $f_j^u = $ 1MHz, $f_k^v=$ 1GHz, $\varsigma_j^u=$ 1.87J/Mb, $\varsigma_k^v=$ 1.87J/Mb, $\varphi_{max}^t=$ 0.99, $\varphi_{max}^e=$ 0.99, $S_j^{max} =$ 20Mbits, $N_k^{max}=$5, $\phi \in$ [0,1], $\psi \in$ [0,1], and $V_j^{max}=$12m/s \cite{jia_learning-based_2021,wang_hybrid_2021}.

To evaluate the performance of VQQ, the optimization results for VQQ, the algorithm Dream from \cite{jia_learning-based_2021} and greedy algorithm are compared in Fig. 3 and Fig. 4. 
It is obvious that the latency and energy cost increase with the increment number of MIoT. 
Besides, compared with the Dream and greedy algorithm, both the latency and energy cost of VQQ are lower. 
It is explained that Dream does not take into account the location change
of UAVs, and the greedy algorithm always offloads the task to the nearest UAVs.
At 30 MIoTs, the latency and energy cost depends mainly on the speed of task transmission.
The VQQ performs slightly better than Dream and significantly better than the greedy algorithm. 
Thus, the effectiveness of the VQQ is proved from Fig. 3 and Fig. 4. Additionally, Fig. 5 and Fig. 6 show the convergence at $I\in \{10,20,30\}$.
With the influence from the environment and the reward from an action, VQQ can quickly converge and gradually become stable.
Thus, the fast convergence reveals the efficiency of VQQ.
\begin{figure}[!t]
  \centering
  \includegraphics[width=8cm]{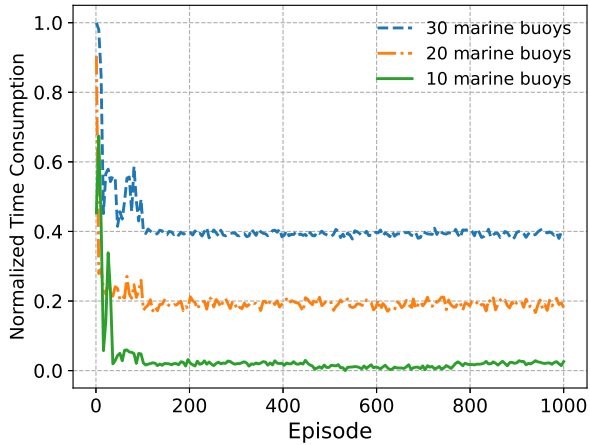}
  \caption{Normalized time consumption \textit{v.s.} episode.}
\end{figure}
\begin{figure}[!t]
  \centering
  \includegraphics[width=8cm]{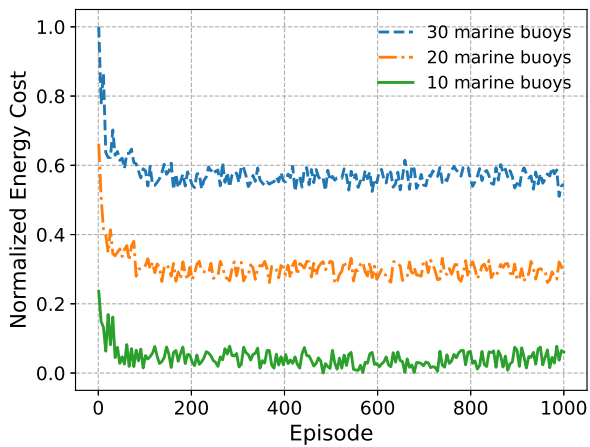}
  \caption{Normalized energy cost \textit{v.s.} episode.}
\end{figure}
\section{CONCLUSIONS}\label{s5}
In this paper, we investigate the cooperation between UAVs and vessels in offloading decisions for maritime communications.
Then, we focus on minimizing the total execution time and energy cost.
However, the problem is in the form of mixed-integer programming, and NP hard to solve.
Hence, to optimize the problem of offloading under uncertain tasks.
We decouple the long-term constraints and use them as a reward for Q-learning.
Then, we design the algorithm of VQQ to efficiently obtain the optimal solution. 
Finally, the simulation results verify the efficiency and convergence of VQQ by compared with two reference algorithms.

\bibliography{references}

\end{document}